\documentclass{svjour2}                    
\smartqed  
\usepackage{graphicx}
\begin{document}

\title{The beneficial role of random strategies in social and financial systems 
}


\author{Alessio Emanuele Biondo\\ Alessandro Pluchino \\ Andrea Rapisarda} 


\institute{A.E. Biondo \at
              Dipartimento di Economia e Impresa - Universit\'a Catania \\
              \email{a.e.biondo@unict.it}           
           \and
         A. Pluchino  \at
              Dipartimento di Fisica e Astronomia and INFN - Universit\'a di Catania\\
              \email{alessandro.pluchino@ct.infn.it}
               \and
         A. Rapisarda  \at
              Dipartimento di Fisica e Astronomia and INFN - Universit\'a di Catania\\
              \email{andrea.rapisarda@ct.infn.it}
}

\date{Received: 25.09.2012 / Accepted: date}

\maketitle

\begin{abstract}In this paper we focus on the beneficial role of random strategies in social sciences by means of simple mathematical and computational models. We briefly review recent results obtained by two of us in previous contributions for the case of the Peter principle and the efficiency of a Parliament. Then, we develop a new application of random strategies to the case of financial trading and discuss in detail our findings about forecasts of markets dynamics.
\keywords{Random strategies \and Sociophysics \and Efficiency \and Numerical simulations \and Peter Principle \and Parliament \and Financial Strategy \and Financial Markets \and Expectations \and Momentum \and RSI}

\end{abstract}

\section{Introduction}
\label{intro}

In science and in physics in particular, noise and randomness are usually restricted to be as low as possible in order to avoid any   influence on the phenomena under examination. Actually, this is not often  possible and one has to live with noise, but, on the other hand, random noise is not always so annoying as one might think intuitively. In fact there are many examples where randomness  has  been  proven to be extremely useful and beneficial. The use of random numbers in science is very well known and Monte Carlo methods are very much used since long time\cite{Binder}, but also in real  physical experiments  random noise has been proven to  be very useful and crucial to explain or help the dynamics: stochastic resonance is probably one of  the most famous and well studied examples of this \cite{Parisi,Gammaitoni}. Of course there are many other cases which support this claim, 
like noice-induced stabilization  \cite{Mantegna}, noise-improved efficiency in communication networks\cite{Caruso}, noise-induced phase transitions \cite{Toral}, etc.
On the other hand, in the last years,  there has been an increasing interest in social phenomena by the physics community \cite{Helbing,Santo-rev,Rapisarda1,Rapisarda2,Buchanan}. Models to study collective behaviors in socio-economical systems, elections mechanisms, consensus formation, management of organizations, spreading of ideas in social networks or herding effects in financial markets, have been proposed and studied with success, providing rigorous and quantitative ways to investigate and help understanding common dynamical laws behind social phenomena. In this respect, it also seems to grow the feeling that the power of  traditional optimization approaches for  solving complex social or economical problems is overestimated, while, on the other hand, it is usually underestimated the role of chance and fluctuations in these fields \cite{mandelbrot,mandelbrot2,mantegna-stanley,Sornette,Taleb}.   
\\
It is just in this spirit that we started, a few years ago, to investigate the possible use of randomness in mathematical and computational models devised to analyze social phenomena. The first application we studied was related to the problems raised by the so-called Peter Principle \cite{Peter,Pluchino1,Pluchino2}. By means of agent based simulations, we demonstrated that random promotion strategies could stop the diffusion of incompetence in hierarchical groups, obtaining also an increase in the global efficiency of the organization under study. Excited by the success of these first studies, we have recently investigated a way to improve  the efficiency of an institution like the  Parliament by means of random selection of  part of its members \cite{Pluchino3}. We present  a rapid overview of the main results obtained for these two applications in sections 2. But the central topic we want to address in this paper concerns the financial market dynamics.    
\\
The very peculiar characteristic of economic and financial systems is that their dynamics depends on their past: economic decisions taken today rely on past expectations, whereas natural laws remain unchanged, no matter what humans think. Thus, economic systems can be considered as feedback-influenced systems, since agents' expectations will influence the entire future dynamics. Such an argument inspired the contribution of many authors who tried to build a mechanism of beliefs formation. We can roughly say that two main reference models of expectations have been widely established within economic theory: the adaptive expectations model and the rational expectation model. We will not proceed in the formal description of such approaches, but we can fruitfully report the main difference between them. The adaptive expectations model (named after Arrow and Nerlove \cite{arrow-nerlove}), has been developed in contributions by Friedman \cite{friedman,friedman2}, Phelps \cite{phelps}, and Cagan \cite{cagan} and assumes that the value of a variable is a somehow weighted average of its past values. Whereas, the rational expectations approach (whose birth dates back to contributions by Muth \cite{muth}, Lucas \cite{lucas}, and Sargent-Wallace \cite{sargent-wallace}, assumes that agents know exactly the entire model describing the economic system and, since they are endowed by perfect information, their forecast for any variable coincides with the objective prediction provided by theory. \\
As one can easily understand, the possibility to make predictions is absolutely central in economic theory. In financial markets, this problem is even more urgent, since the extremely high volatility and the strong instability that everyday can be observed on those markets. This leads to articulated theories of trading that try to forecast market dynamics in order to realize profits from intermediation. There is not unanimity of opinions, in economic literature, about the actual possibility of traders to predict financial values. The so-called Efficient Market Hypothesis, which refers to the rational expectation models, consider the role of rationality of agents as the main and most important part of market dynamics, whereas an adaptive approach is oriented in building forecasts from past dynamics.        \\ 
Financial crises showed that mechanisms and trading strategies are not immune from failures. Their periodic success is not \emph{free of charge}: catastrophic events burns enormous values in dollars. Are we sure that elaborated strategies fit the unpredictable dynamics of markets? Are analysts aware that without having complete information, with non perfect markets, no rational mechanism can be invoked in financial transactions? 
In order to reply to these questions a simple simulation will be done. It will perform a comparative analysis of performances between trading strategies: two very famous and reliable technical strategies (namely \emph{momentum} and \emph{RSI-divergence} \cite{wilder,murphy}) that traders adopt everyday for their operations in real markets, \emph{versus} a random strategy. Rational expectations theorists would immediately bet that the random strategy will easily loose the competition, but this is not the case, as we discuss in section 3, where we present new detailed numerical results. Conclusions are drawn in Section 4.

\section{Improving the efficiency of social groups or organizations by means of random strategies }

\subsection{The case of the Peter Principle}

The  Peter principle  was enunciated by the Canadian psychologist Lawrence J. Peter in a famous book of the $60$s \cite{Peter} based on some sensible assumptions on the transfer of skills from one level of a hierarchical organization to the next one. The principle states that {\it "in a hierarchical organization each member rises the hierarchy, as a result of meritocratic promotions, up to when he/she reaches its minimum level of competence".} Even if it sounds paradoxical, according to Peter such a perverse effect surely occurs whenever promoted people do change their task passing from the previous to the next level: in this way, incompetence will inevitably spread at the top of the organization, endangering its proper functioning. In refs. \cite{Pluchino1,Pluchino2} it has been demonstrated by means of numerical simulations, that the principle is true under certain conditions and that one  can overcome its effects  by adopting random promotions. In the following we explain the details of the models used and the main results obtained. 
\\
The first model studied  in ref. \cite{Pluchino1},  considered  a  schematic pyramidal organization  with  160 positions divided into six levels. Each level had a different number of members  with a different responsibility according to the hierarchical position. The members of the organization  were characterized by their age, in the interval 18-60 years, and their degree of competence in the range 1-10. As initial conditions we selected ages and competences following normal distributions. 
At each time step of one year, if members reach  an age over the retirement threshold ( fixed at  60 years) or have  a competence lower than the dismissal threshold (fixed at 4) they leave the organization  and someone from the level immediately below (or from outside for the initial bottom level) has to be selected  for promotion. Four different competing strategies of promotions were taken into account: promotion of the best worker, promotion of the worst, promotion of  a random worker and promotion in an alternate way  of  the best and of   the worst. 
Two different mechanisms of competence transmission were also considered:
\\
1) �Common Sense (CS)� - if the features required from one level to the upper are enough stable, the new competence at the upper level is correlated with the previous one and the agent maintains his competence with a small error;
\\
2) �Peter Hypothesis (PH)� - if the features required from one level to the upper can change considerably, the new competence at the upper level is {\it not correlated} with the previous one, so the new competence is again  randomly assigned  for each promotion.
\\
The global efficiency $E$ was defined by summing the competences of the members level by level, multiplied by the level-dependent factor of responsibility, ranging from 0 to 1 and linearly increasing on climbing the hierarchy. 
If $C_i$ is the total competence of level $i$-th, the global efficiency can be written as
$
 E(\%)=\frac{\sum_{i=1}^6 C_i r_i}{Max(E) \cdot N} \cdot 100  ,
$
where  $Max(E)=\sum_{i=1}^6 (10 \cdot n_i) \cdot r_i / N$, and  $n_i$ is the number of agents of level $i$-th.
\\
The main results found after averaging  over many different realizations of the initial conditions, confirmed the risk of incompetence spreading when  the Peter Hypothesis holds.  In particular it was  found that  promoting the best members is a winning  strategy only if the CS hypothesis holds, otherwise is a loosing one. On the contrary, if the PH holds, the best strategy becomes that one of even promoting the worst member.
On the other hand, if one {\it does not know} which of the two hypothesis holds, then adopting a random promotion strategy, or alternating the promotion of the best and the worst candidates, results to be always a winning choice.
\\
Although the paper was quite successful and appreciated also for its simplicity \cite{ignobel}, its  paradoxical results needed a confirmation within a more realistic model.   
\begin{figure}
\begin{center}
 \includegraphics[width=0.8\textwidth]{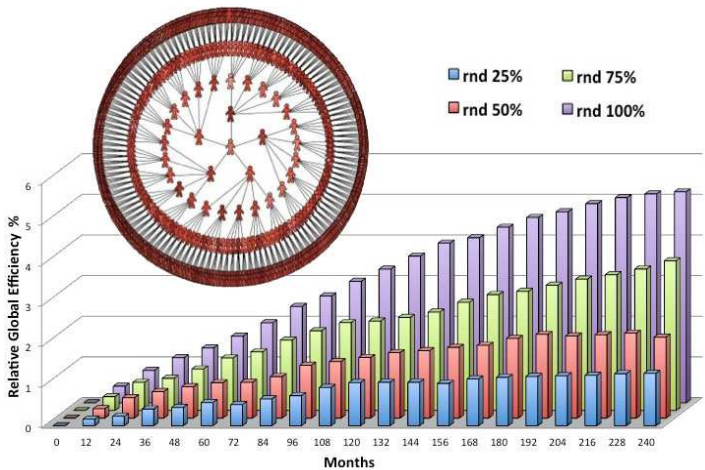}
\caption{The  dynamical evolution of the model efficiency gain  during the first 20 years. One  can observe an immediate increase of the efficiency  since the beginning of the adoption of a random strategy with respect to a meritocratic one with the Peter hypothesis, see text for further details. The topology of the modular hierarchical organization considered is also shown in the top left part of the figure.}
\label{fig:1}      
\end{center}
\end{figure}
To this end in a second model \cite{Pluchino2} it was adopted a schematic modular organization, i.e. a hierarchical tree network with $K=5$ levels, where each agent (node) at levels $k=1,2,3,4$ (excluding the bottom level with $k=5$) has exactly $L$ first subordinates (i.e. first neighbors at level $k+1$), which will fill that position when it will become empty.  
This means that, at variance with the pyramidal schematic model of our first paper, in this case promoted agents could follow the links to ascend through levels. On the other hand, neglecting the links and promoting agents from the entire level $k$ to the next level $k-1$, one could recover the pyramidal model as a particular case. 
\\
In the inset of Fig.1 we show an example of such a hierarchical tree network, with $K=5$ levels and $L=5$, for a total of $N=781$ agents. The responsibility value is $0.2$ for the bottom level and increases linearly, like in the previous model, with step $0.2$ for each level up to the top one, whose responsibility value is $1$.
Another improvement of the second model was the time units adopted, which becomes one month instead of one year. 
Moreover, instead of studying the improvements with respect to an arbitrary initial state for the organization (with an arbitrary value of the initial global efficiency, as done in \cite{Peter}), we evaluated a {\it relative global efficiency} $E_r(\%)$, calculated with respect to a fixed transient during which a meritocratic strategy (i.e. the promotion of the best workers coupled with the Peter hypothesis) was always applied.  We introduced also a new "mixed" strategy, where a different increasing percentage of random promotions with respect to meritocratic strategy is considered. 
Finally, we either considered promotions of a member from a level to the next one following the links ({\it neighbors mode}) or without considering the links of the hierarchical tree ({\it global mode}), in order to reduce the new model to that one considered in ref. \cite{Pluchino1}. 
\\
In Fig.1 we report, as an example, the relative efficiency as function of time for the {\it neighbors mode}. 
A period of $240$ months ($20$ years) was considered and an average over $30$ different realizations of the initial conditions was performed in order to diminish the effect of fluctuations. Our aim was to investigate the effects of the introduction of an increasing percentage of random promotions (from $25\%$ to $100\%$) within an otherwise meritocratic strategy, under the Peter hypothesis of competence transmission. The plot clearly shows how even a moderate amount of randomness increases the efficiency of the organization in a rapid and substantial way. This second model thus confirms the results of the previous one, even for larger organizations and different topologies: random strategies provide a good advantage, in terms of efficiency, with respect to a full "naively" meritocratic system of promotions and, at the same time, diminish the risk of the rising of incompetence to the top of the organization. One can refer to ref.\cite{Pluchino2} for more details.

\subsection{The case of the Parliament}

Stimulated by the results about the Peter principle, we started to ask if random strategies could be useful also in the selection of members of political institutions. In this section we discuss a recent application of random strategies for improving the efficiency of a prototypical Parliament and present the main results obtained by two of us in ref.\cite{Pluchino3}. 
Inspired by the so called 'Cipolla's diagram' \cite{Cipolla}, we realized a virtual model of one chamber of a Parliament by characterizing its members through their attitude to promote personal and general interest through legislative proposals. In this way we represented individual legislators as points in the two dimensional Cipolla's diagram, where on the $x$ axis is reported the personal gain and on the $y$-axis the social gain (considered as the final outcome of trading relations produced by the laws). Both the $x$ and $y$ coordinates of each legislator are real numbers included in the interval $[-1,1]$.
In our simulations we considered a Parliament with $N$ members and two parties or coalitions, $P_1$ (the majority one) and $P_2$ (the minority one), with a different percentage of members. All the points representing the members of a party will lie inside a circle, with a center whose position on the Cipolla's diagram is fixed by the average collective behavior of all its members, and with a given radius $r$ that fixes the extent to which the Party tolerates dissent within it (the larger this radius, the greater the degree of tolerance within the party; for this reason this circle was called  {\it circle of tolerance} of the party).
\\
In \cite{Pluchino3} we found that the efficiency of the Parliament, defined as the product of the percentage of the accepted proposals times their overall social welfare, can be influenced  by the introduction of a given number $N_{ind}$ of {\it not elected but randomly selected} legislators, called 'independent' since we assume that they remain free from the influence of any party. These independent legislators will be represented as free points on the Cipolla diagram.
The dynamics of our model is very simple. During a legislature $L$ each legislator, independent or belonging to a party, can perform only two actions: 
\\
(i) he/she can propose one or more acts of Parliament, with a given personal and social advantage depending on his/her position on the diagram. 
\\
(ii) he/she has to vote for or against a given proposal, depending on his/her {\it acceptance window}, i.e. a rectangular subset of the Cipolla diagram into which a proposed act has to fall in order to be accepted by the voter (whose position fixes the lower left corner of the window). The main point is that, while each free legislator has his/her own acceptance window, so that his/her vote is independent from the others� vote, all the legislators belonging to a party always vote by using {\it the same} acceptance window, whose lower left corner corresponds to the center of the circle of tolerance of their party. Furthermore, following the party discipline, any member of a party accepts {\it all} the proposals coming from any another member of the same party (see \cite{Pluchino3} for further details). 
\begin{figure}
\begin{center}
 \includegraphics[width=1.1\textwidth]{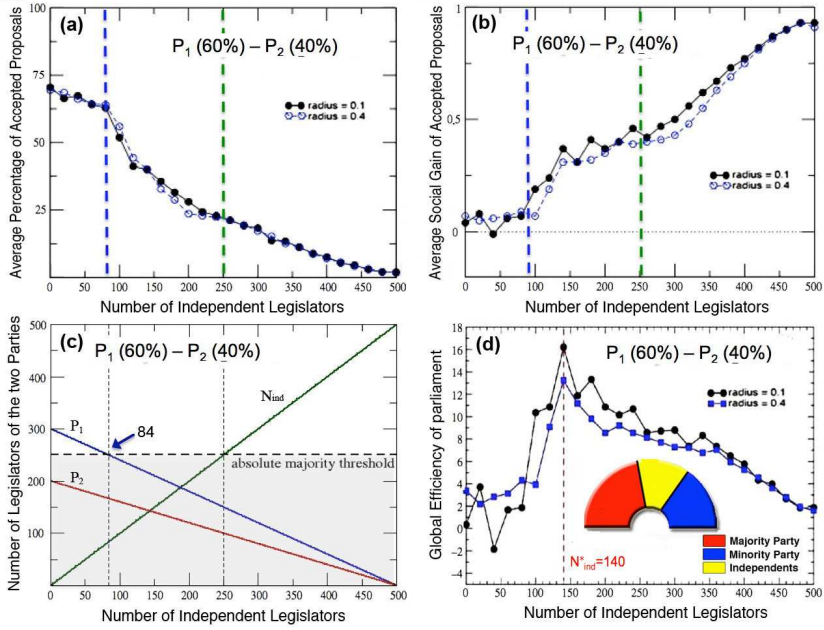}
\caption{Simulation results for a Parliament with $N=500$ members, two parties $P_1$ and $P_2$ and $N_{ind}$ independent (randomly selected) legislators. Panel (a): Average percentage of accepted proposals vs $N_{ind}$; Panel (b): Average overall social gain vs $N_{ind}$; Panel (c): Size of the three Parliament components ($P_1$, $P_2$ and $N_{ind}$) vs $N_{ind}$; Panel (d): Global efficiency vs $N_{ind}$. All the numerical points represent averages over $100$ different legislatures. See text. }
\label{fig:2}       
\end{center}
\end{figure}
Once {\it all} the $N$ members of Parliament voted for or against a certain proposal, the latter will be accepted only if receives at least ${N\over2}+1$ favorable votes. At this point we can calculate the efficiency $Eff(L)$ of the Parliament during a legislature $L$ by simply multiplying the percentage of accepted proposals $N_{acc}(L)$ times the overall social gain $Y(L)$ they ensure (notice that $Eff(L)$ will be therefore expressed by a real number included in the interval [-100,100]). In this respect, we investigated how the three quantities $N_{acc}(L)$, $Y(L)$ and $Eff(L)$ change as function of the number $N_{ind}$ of independent legislators introduced in the Parliament.   
\\
In Fig.2 we present some of the main results of \cite{Pluchino3}, obtained by simulating a Parliament with $N=500$ members and two parties $P_1$ and $P_2$ with - respectively - $60\%$ and $40\%$ of legislators. Notice that these latter values represent the percentages of seats assigned to both the majority and minority parties in a given legislature {\it after} having reserved the $N_{ind}$ seats to the independent legislators, therefore are values that decrease by increasing $N_{ind}$.  
\\
In panels (a) and (b) we plot, respectively, the percentage of accepted proposals and the correspondent social gain as function of the number of independent legislators, averaged over a set of $N_L=100$ legislatures, each one with a total number of $1000$ proposals but with a different random distribution of legislators and parties on the Cipolla's diagram. We also repeated all the simulations for two different values of the radius $r$ of both the parties ($0.1$ and $0.4$). 
It clearly appears that, in average, the introduction of an increasing number of independent legislators causes a decrease in the percentage of accepted acts (since reduces the weight of the party discipline in accepting each proposal) but, simultaneously, also produces an increase of the average social gain of the same accepted acts (since only the proposals ensuring a higher social gain succeed in being accepted by the majority of the Parliament in presence of independent legislators). In both the curves two different threshold values of $N_{ind}$, corresponding to a change in the slope, can be recognized and can be easily explained looking to panel (c), where the size of the two parties $P_1$ and $P_2$ are reported as function of $N_{ind}$ (this point was not discussed in \cite{Pluchino3} so it still needs a clarification). It results that (for our choice of parameters) the party $P_1$ looses the absolute majority in the Parliament for $N_{ind}>84$, therefore only over this first threshold $N_{acc}$ and $Y(L)$ start to significantly change. The second threshold, on the other hand, takes place when the independent component becomes, in turn, the absolute majority of legislators, thus accelerating - respectively - the decreasing and increasing trends of $N_{acc}$ and $Y(L)$. In any case, despite of these explanation, it remains absolutely not trivial to predict the exact shape of these non linear curves, and this will reflect on the difficulty of {\it a-priori} determining the resulting efficiency.       
\\
Finally, in panel (d), we plot the product of the two previous quantities, therefore obtaining ({\it a-posteriori}) the global efficiency of the Parliament (averaged over the $100$ legislatures). It is worthwhile to notice here that:
\\
(i) for any value of $N_{ind}$ the global efficiency shows an increment with respect to the two extreme cases $N_{ind}=0$ (only parties) and $N_{ind}=N$ (only independent members and no parties); this means that, in analogy with the results shown in the previous subsection, even a small degree of randomness added to the system is able to increase its performance;
\\
(ii) the combination of the two previous curves gives rise to a pronounced peak in efficiency in correspondence of a critical value $N_{ind}^*=140$ of independent legislators, which does not change with the radius $r$ but only depends on the relative size of the two parties. In \cite{Pluchino3} we discovered that it is possible to write down an analytical formula, called {\it efficiency golden rule}, able to exactly predict the value $N_{ind}^*$ as function of the total number of legislators $N$ and the size $p$ (in percentage) of the majority party $P_1$. The formula is the following: $N_{ind}^*= \frac{2N - 4N\cdot(p/100) + 4}{1 - 4\cdot(p/100)}$. It allows us to imagine a new electoral system where, after ordinary elections for determining the relative sizes of the majority and minority parties, one could use our 'golden rule' to find out the optimal number of independent legislators, chosen at random among all the citizens willing to candidate (out of the parties system), able to maximize the Parliament efficiency. 
\\
In conclusion, in this section we have shown a couple of applications of numerical simulations to management and politics, with the aim to convince the reader that some degree of randomness could play a constructive role in improving the efficiency of our institutions. In the following section we will give further support to this hypothesis presenting a new application of random strategies to financial markets.

\section{Financial Markets, Randomness and Trading Strategies: The Case Study of FTSE UK All-Share index}   

In $2001$ a valued british psychologist, Richard Wiseman, performed an eccentric experiment in order to test the predictive power of the trading strategies in the financial markets \cite{wiseman}. He gave the same virtual amount of money ($5000$ pounds) to three very different people, a London's financial trader, an astrologer and a four years old child named Tia, asking them to invest the money in the UK stock market (the trader following his algorithms, the astrologer following the stars' movements, and Tia completely at random). At the end of a very turbulent year for the world financial markets, the result of the competition was completely unexpected: if, on one hand, the trader and the astrologer had \textit{lost} respectively $46.2\%$ and $6.2\%$, on the other hand, with the help of her random strategy, Tia had even \textit{earned} $5.8\%$! Other similar experiments, with similar results, were performed by substituting the child with a chimpanzee \cite{wiseman} or by selecting securities by darts \cite{porter}. However, as far as we know, no one has yet tried to test the effectiveness of random strategies in finance through computer simulations. This is exactly what we will do in this section, by using the FTSE UK All-Share index series of the last $14$ years as case study.
\begin{figure}
\begin{center}
\includegraphics[width=0.9\textwidth]{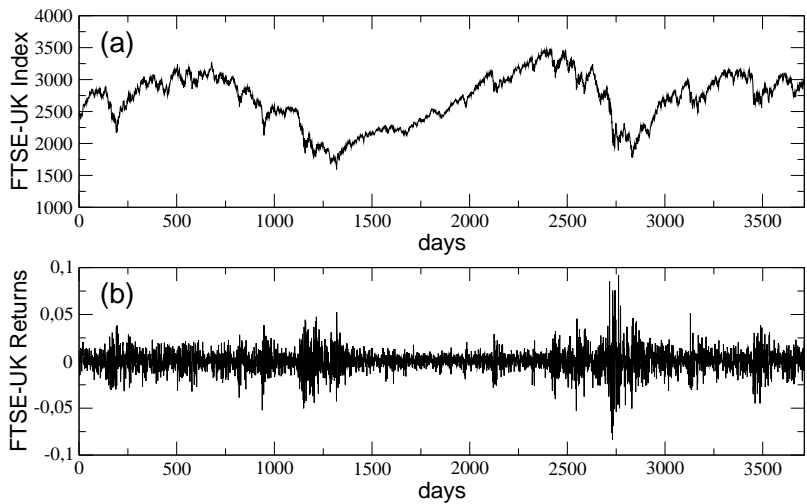}
\caption{Panel (a): Behavior of the FTSE UK All-Share index from January, 1st 1998 to August, 3rd 2012 (14 years, 3714 days). Panel (b): Returns series for the FTSE UK All-Share index in the same period.}
\label{fig:3}       
\end{center}
\end{figure}
\\
In panel (a) of Fig.3 we plot the behavior of the FTSE UK All-Share index $I(t)$ from January, 1st 1998 to August, 3rd 2012, for a total of $T_{UK}=3714$ days, while in the panel (b) we report the correspondent 'returns series', calculated as the ratio 
$[I(t+1)-I(t)]/I(t)$ (as usually defined in the econophysics literature, see e.g. \cite{mantegna-stanley} p.38). 
From the latter it immediately appears, by imagining to divide the time series into three trading windows of equal size (of around $1200$ days each), that the index behavior alternates a first intermittent period with a more regular one, ending again with a last intermittent interval. A finer resolution would reveal a further, self-similar, alternation of intermittent and regular behavior over smaller time scales, a well known feature (which also resembles turbulence phenomena) characterizing financial markets \cite{mandelbrot,mandelbrot2,mantegna-stanley}. As previously anticipated, our goal is to test the performance of three different trading strategies (each hypothetically corresponging to one trader) in simply predicting, day by day, the upward ('bullish') or downward ('bearish') movement of the index $I(t+1)$ at a given day with respect to the closing value $I(t)$ one day before: if the prediction is correct, the trader wins, otherwise he/she looses. In this respect we are only interested, here, in evaluating the percentage of wins or losses guaranteed by each strategy and at different time scales, assuming that - at every time step - the traders perfectly know the past history of the FTSE UK All-Share index but do not possess any other information and cannot neither exert nor receive any influence to or from the market. 
\\
The reader must be aware that we want to evaluate the  individual investor's \textit{ex-ante} predictive capacity without assuming \textit{perfect arbitrage} in the context of a theoretical general equilibrium approach to financial markets. This is particularly important in order to avoid misleading interpretations that could consider the previously described set of information given to our traders as the weak form of Efficient Markets Hypothesis paradigm. The difference is both theoretical and methodological: after Fama \cite{fama} we define efficiency in financial markets according to the existence of perfect arbitrage. This implies that if markets were inefficient, there would exist the possibility of unexploited profits and traders would immediately operate to obtain highest remunerations. Jensen \cite{jensen} confirms such an implication exactly combining the informative set available and the existence of trading: only when it is impossible to make further profits, given an informative set, financial markets are efficient. This is the rationale behind the distinction made by Fama \cite{fama} among three degrees of efficiency: namely "weak", "semi-strong", and "strong", according with the completeness of the informative set. Thus, it is theoretically consequent that, if the Efficient Markets Hypothesis held, the financial markets would result complete, efficient and perfectly competitive. This implies that, in presence of complete information, randomness should play no role, since the Efficient Market Hypothesis would generate a perfect trading strategy, able to predict exactly the market values, embedding all the information about short and long positions worldwide. On the other hand, the lack of complete information makes efficiency impossible to be reached. In this case, randomness could be a \textit{beneficial} tool to face the lack of information.
We test here precisely this hypothesis: if a trader presumed the lack of complete information through all the market (i.e. the unpredictability of stock prices dynamics \cite{bachelier,samuelson,blackscholes,merton,coxingersollross,hullwhite,heston}), an \textit{ex-ante} random trading strategy would perform, on average, as good as well-known trading strategies. 
\\
The three strategies we will adopt in the present study are the following:
\\
{\it 1) Random (RND) Strategy}
This  strategy is the simplest one, since the correspondent trader makes his/her 'bullish or 'bearish' prediction at the time $t$ completely at random (with uniform distribution), like Tia in the Wiseman's experiment. 
The other two strategies, on the contrary, are based on two indicators that are very well known by financial traders. 
\\
{\it 2) Momentum-based (MOM) Strategy}
This strategy is based on  the so called 'momentum' $M(t)$ indicator, i.e. the difference between the value $I(t)$ and the value $I(t-\tau_M)$, being $\tau_M$ a given trading interval (expressed in days). Then if $M(t)=I(t)-I(t-\tau_M)>0$, the trader predicts an increment of the closing index for the next day (i.e. it predicts that $I(t+1)-I(t)>0$) and vice-versa. In the following simulations we will consider $\tau_M=7$ days, since this is one of the most used time lag for the momentum indicator. 
\\
{\it 3) RSI-based Strategy}
This latter strategy is based on a more complex indicator  called 'RSI' (Relative Strength Index) \cite{wilder}. It is considered a measure of the stock's recent trading strength and its definition is: $RSI(t)=100-100/[1+RS(t)]$, where $RS(t,\tau_{RSI})$ is the ratio between the sum of the positive returns and the sum of the negative returns occurred during the last $\tau_{RSI}$ days before $t$. Once calculated the RSI index for all the days included in a given time-window of length $T_{RSI}$ immediately preceding the time $t$, the trader which follows the RSI strategy makes his/her prediction on the basis of a possible reversal of the market trend, revealed by the so called 'divergence' between the original FTSE UK All-Share series and the new RSI one (see \cite{wilder} for more details). In our simplified model, the presence of such a divergence translates into a change in the prediction of the $I(t+1)-I(t)$ sign, depending on the 'bullish or 'bearish' trend of the previous $T_{RSI}$ days. In the following simulations we will choose $\tau_{RSI}=T_{RSI}=14$ days, since - again - this value is one of the mostly used in RSI-based actual trading strategies \cite{murphy}.   
\begin{figure}
\begin{center}
\includegraphics[width=1\textwidth]{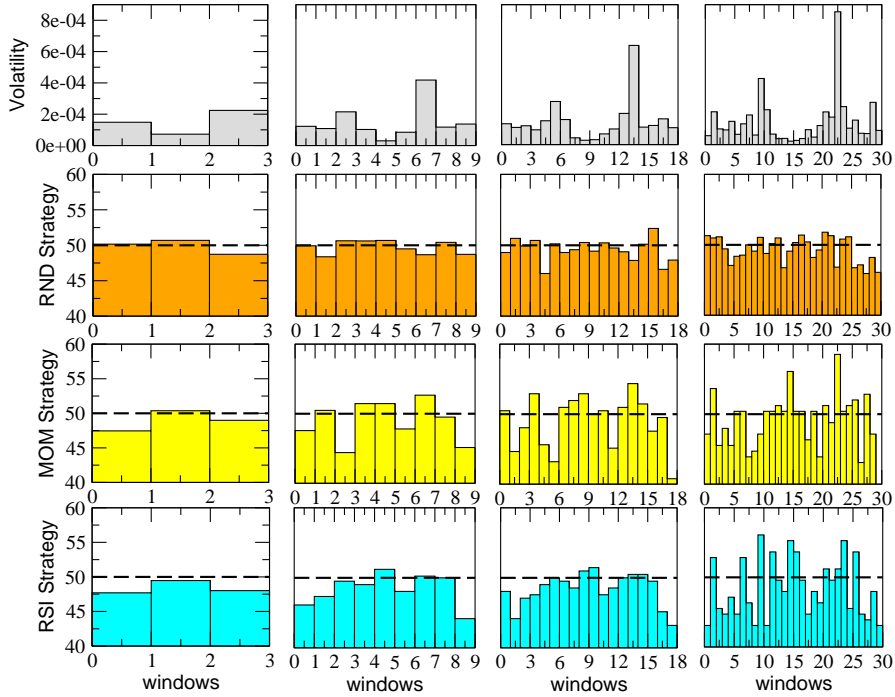}
\caption{Simulations results: the FTSE UK All-Share series is divided into an increasing number of trading-windows of equal size, in order to simulate different time scales. In the first row the volatility of the index, calculated inside each window, is shown for comparison. In the other rows, the percentages of wins for the three strategies, averaged over $10$ different runs inside each window, are reported. A $50\%$ dashed line is also plotted as reference. See text. }
\label{fig:4}       
\end{center}
\end{figure}
\\
In Fig.4 we report a first comparison of the simulations results for our three strategies, applied to the FTSE UK All-Share series. In particular, we test the performance of the strategies by dividing the whole series into a sequence of $N_w$ trading windows of equal size $T_w=T_{UK}/N_w$ (in days) and evaluating the number of wins for each strategy inside each window while the traders move along the series day by day, from $t=0$ to $t=T_{UK}$.
This procedure, varying $N_w$, allows us to explore the behavior of the various strategies at several time scales (ranging, approximatively, from $6$ months to $5$ years).   
In the first row of Fig.4 we plot the volatility of the FTSE UK All-Share index, calculated for $4$ increasing values of  $N_w$: from left to right, we consider $3$, $9$, $18$ and $30$ windows with size $T_w$ equal to, respectively, $1237$, $412$, $206$ and $123$ days. In the three rows below we plot, in correspondence of the same $4$ windows configurations, the values of the percentage of wins for the three strategies   
within each window, averaged over $10$ different runs (in this first set of simulations such an average is meaningful only for the random strategy, since the other two strategies are completely deterministic, once fixed their characteristic parameters and the trading series). 
\begin{figure}
\begin{center}
\includegraphics[width=0.8\textwidth]{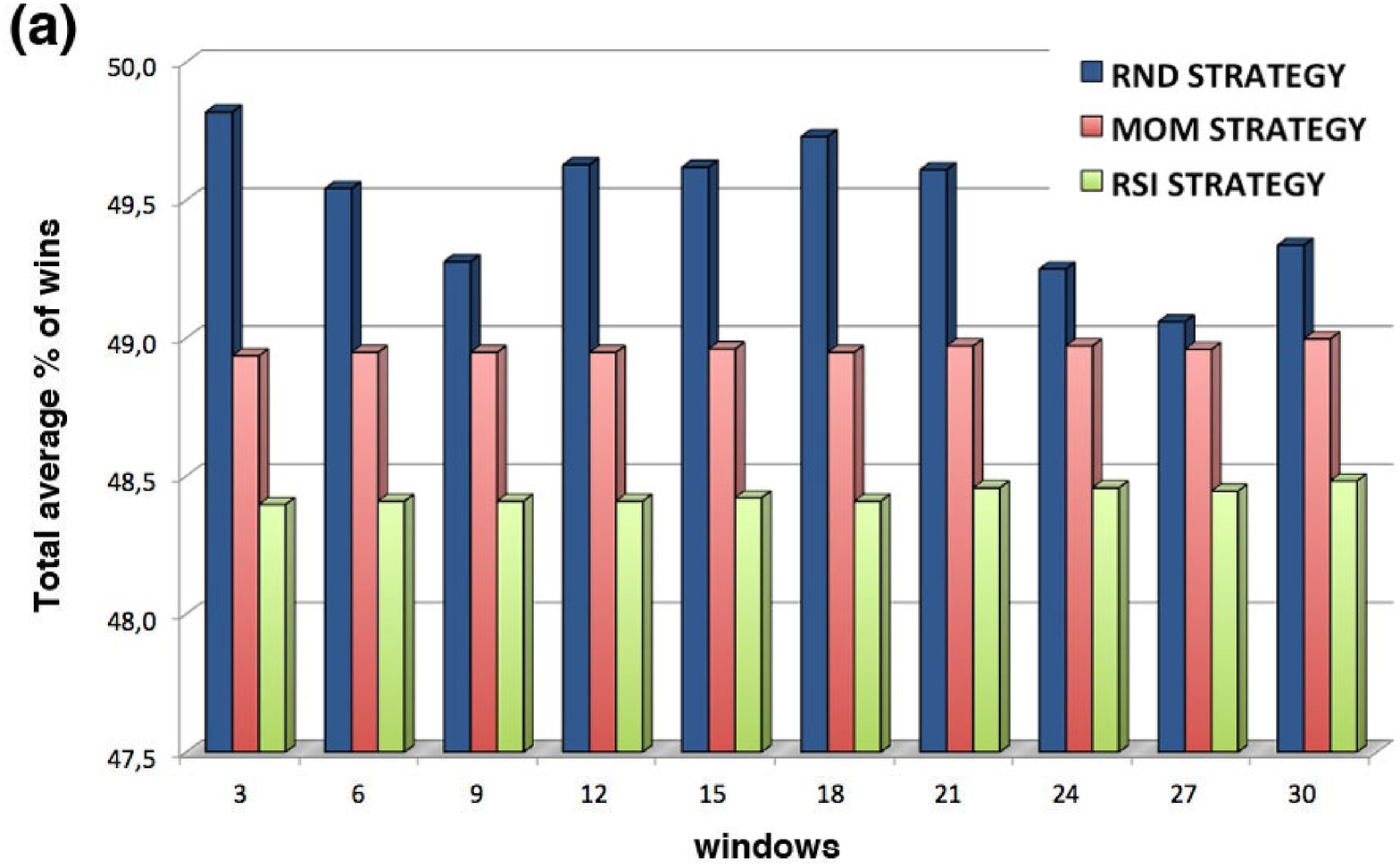}
\includegraphics[width=0.8\textwidth]{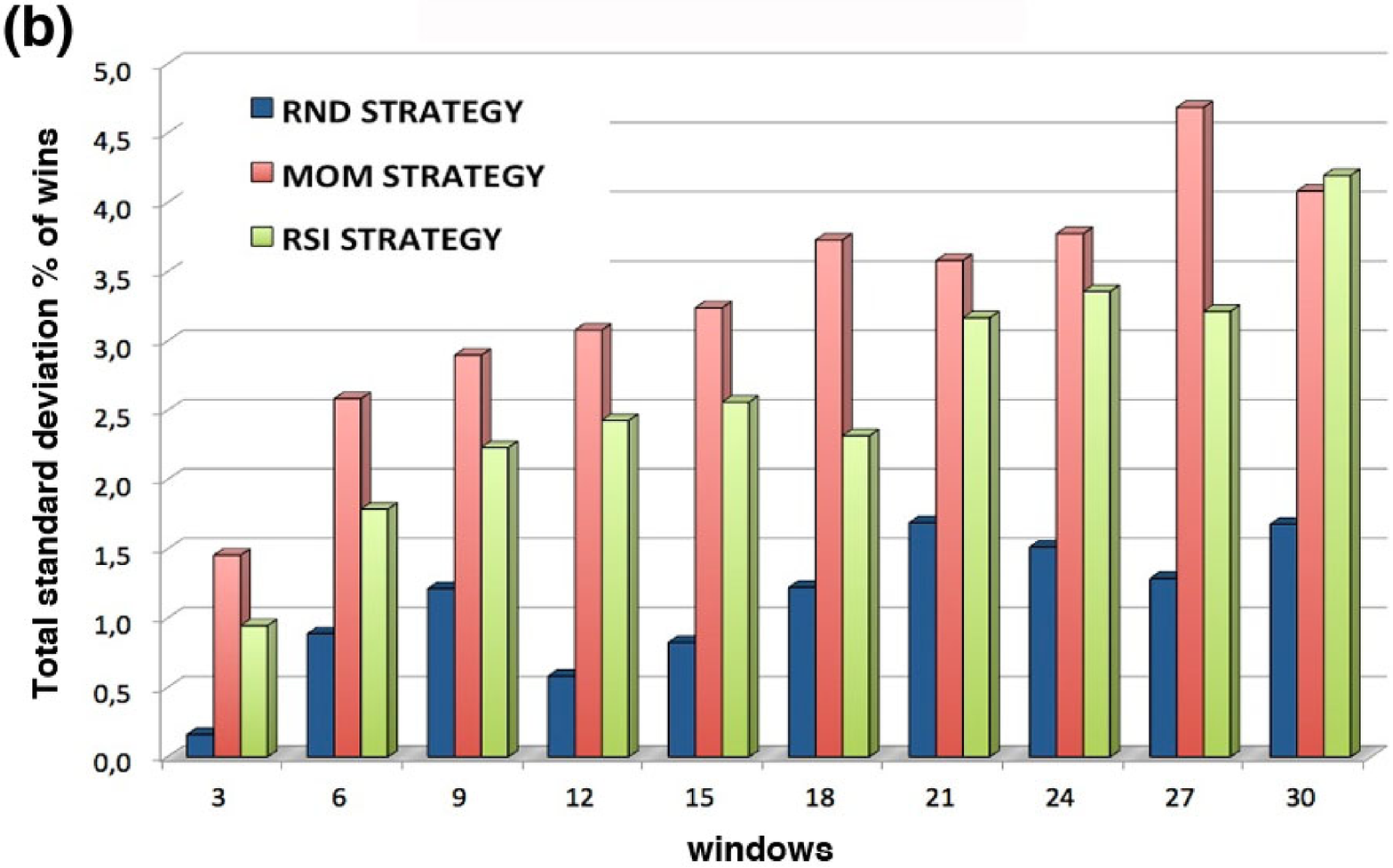}
\caption{Panel (a): Percentages of wins for the three strategies (magnified on the $y$-axis in order to better appreciate the comparison) averaged over {\it all the windows} in each one of $10$ different subdivisions of the FTSE UK All-Share series, with an increasing number of windows (reported on the $x$-axis). Panel (b): The correspondent total standard deviations for the same configurations of windows and for the three strategies. See text.}
\label{fig:5}       
\end{center}
\end{figure}
\\
Differences between the Random  strategy (RND, second row) and the two standard trading strategies (MOM and RSI, third and fourth row) are evident.  
Actually, at any time scale (but in particular for large values of $N_w$), the RND appears much less fluctuating (i.e. less risky) than the others. Furthermore, MOM and RSI performances seem to behave slightly worse than the RND one at the beginning and at the end of the whole FTSE UK All-Share series, i.e. when the market behavior is more intermittent - as shown by the correspondent higher volatility. 
\\
In Fig.5 we can better appreciate this quite surprising result by observing, in panel (a), the percentage of wins for the three strategies averaged over {\it all the windows} in each one of several configurations with different $N_w$ (ranging from $3$ to $30$ with step $3$) and, in panel (b), the correspondent standard deviations. From the first histogram it appears that the average gains of the three strategies are comparable and restricted in a narrow band just below the $50\%$ of wins, with a slight advantage of the RND one (which, however, could depend on the trading series chosen for the analysis). At the same time, the second histogram confirms the higher stability of the random strategy over the other ones. The fact that none of the three strategies overcomes the threshold of $50\%$ could seem paradoxical, but we stress that this is true only averaging over the whole FTSE UK All-Share series, whereas, of course, they can exceed that threshold within single trading windows, as also visible in Fig.4. These findings seem to suggest that - as also observed by Taleb \cite{Taleb} - the success of a trading strategy at a small time scale would probably depend much more on luck than on the real effectiveness of the adopted algorithm, since on a large time scale its performance is comparable with (or, as in this case, even worse than) a random one.
\begin{figure}
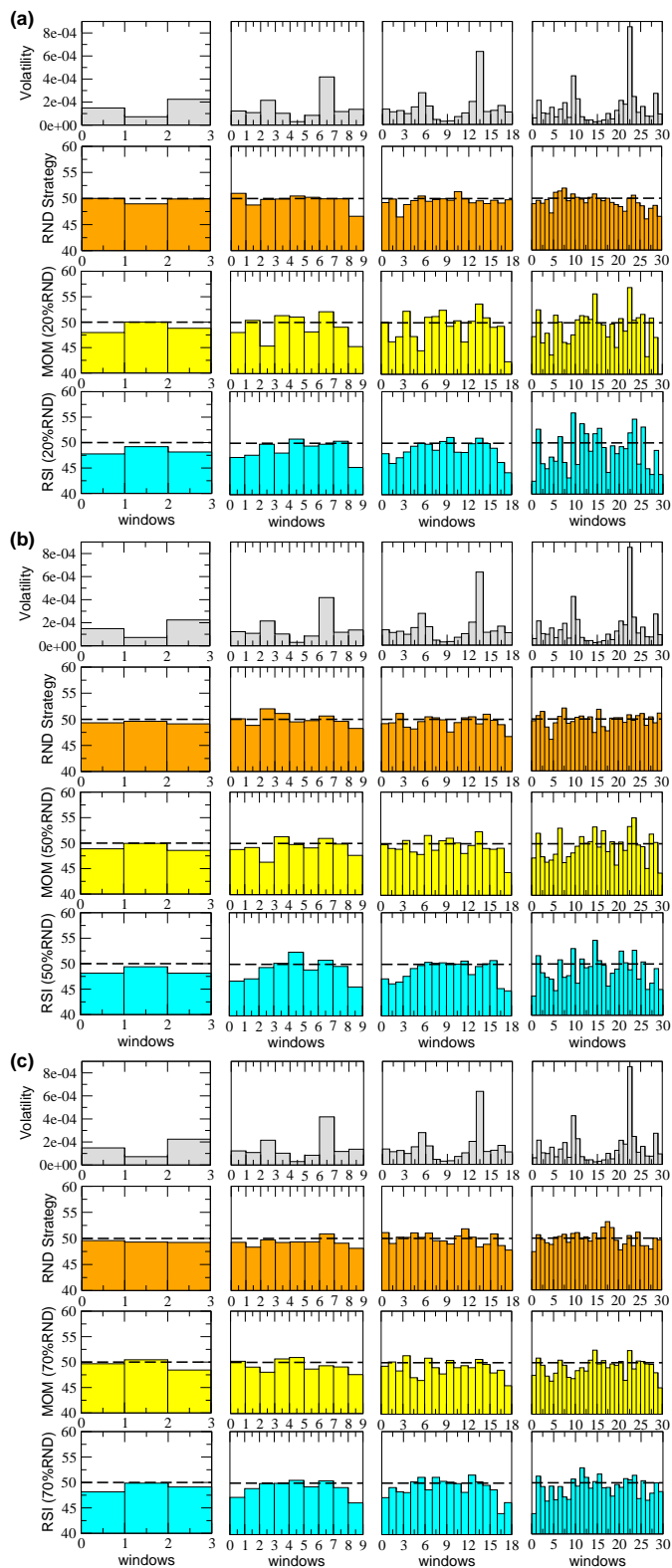

\begin{center}
\includegraphics[width=0.75\textwidth]{figure6a.eps}
\includegraphics[width=0.75\textwidth]{figure6b.eps}
\includegraphics[width=0.75\textwidth]{figure6c.eps}
\caption{Volatility and average percentage of wins (inside each window and over $10$ runs) for the three traders, as in Fig.4, but with an increasing quantity of randomness mixed with the MOM and the RSI strategies. Panel(a): $P_{RND}=20\%$; Panel (b): $P_{RND}=50\%$; Panel (c): $P_{RND}=70\%$. See text. }
\label{fig:6}       
\end{center}
\end{figure}
\\
At this point, following the results surveyed in the previous section, one may suspect that the introduction of some randomness into the standard, otherwise deterministic, trading strategies (MOM and RSI), could play a beneficial role. In the next figures we actually show that this is indeed the case.
\begin{figure}
\begin{center}
\includegraphics[width=0.48\textwidth]{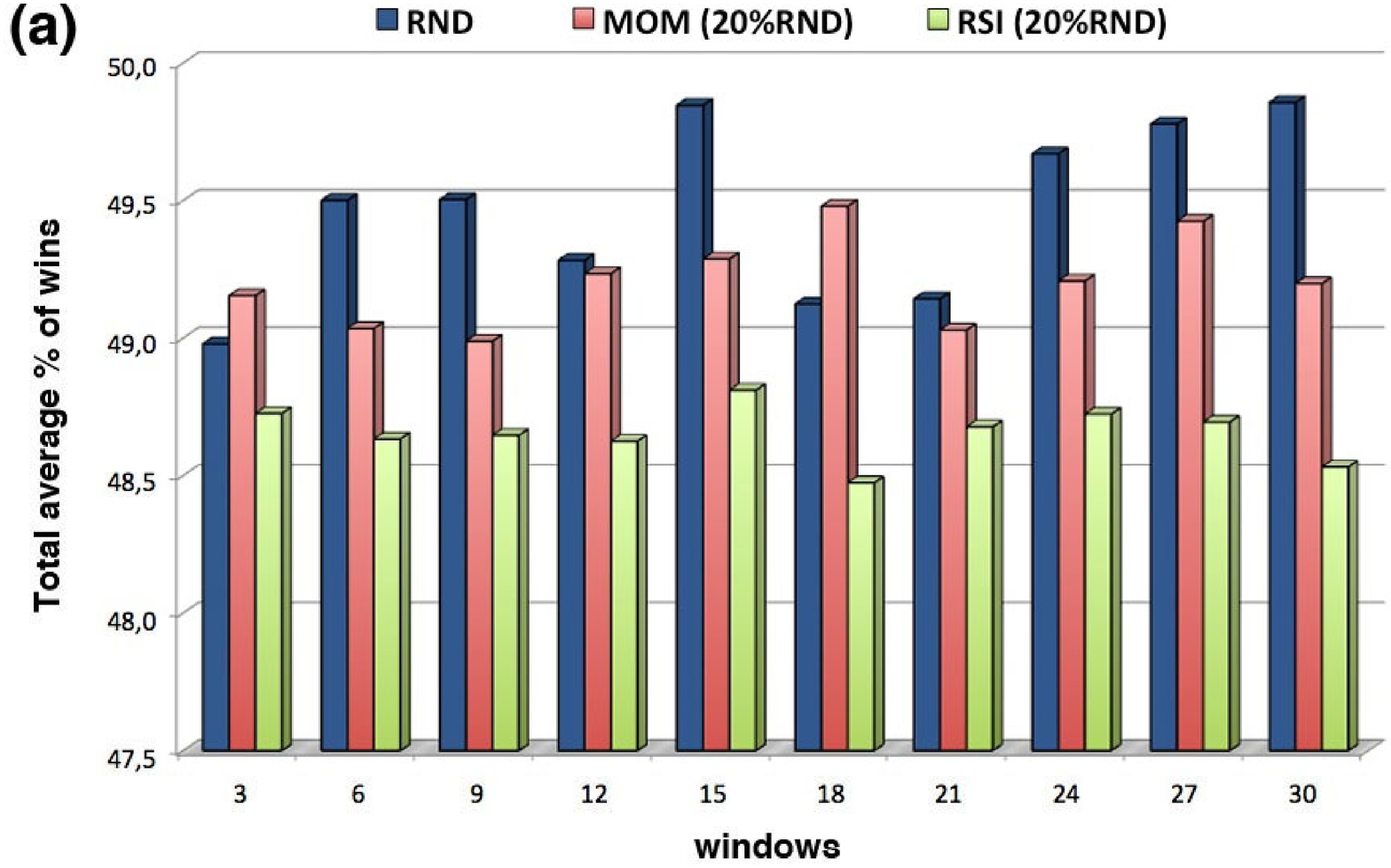}
\includegraphics[width=0.48\textwidth]{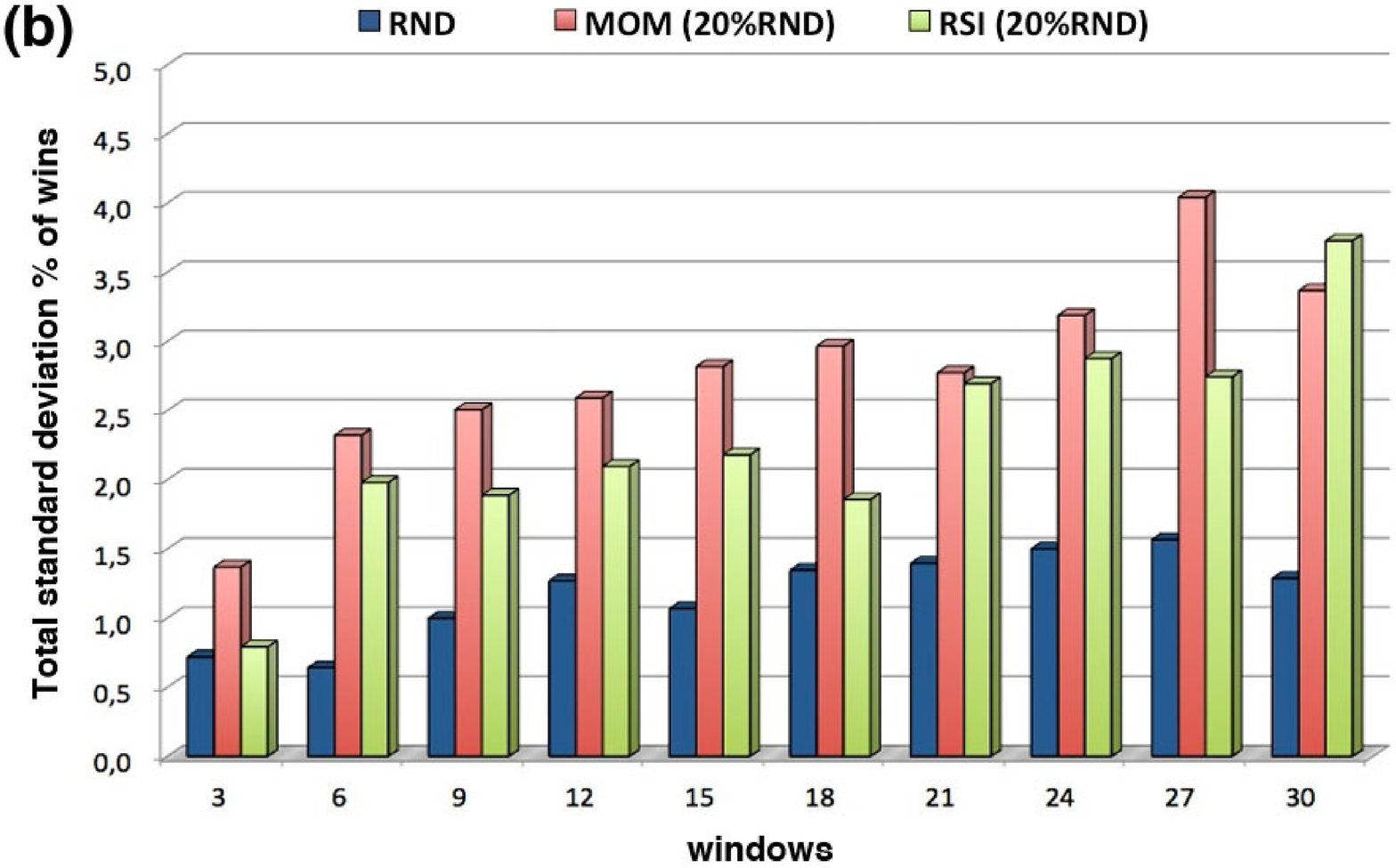}
\includegraphics[width=0.48\textwidth]{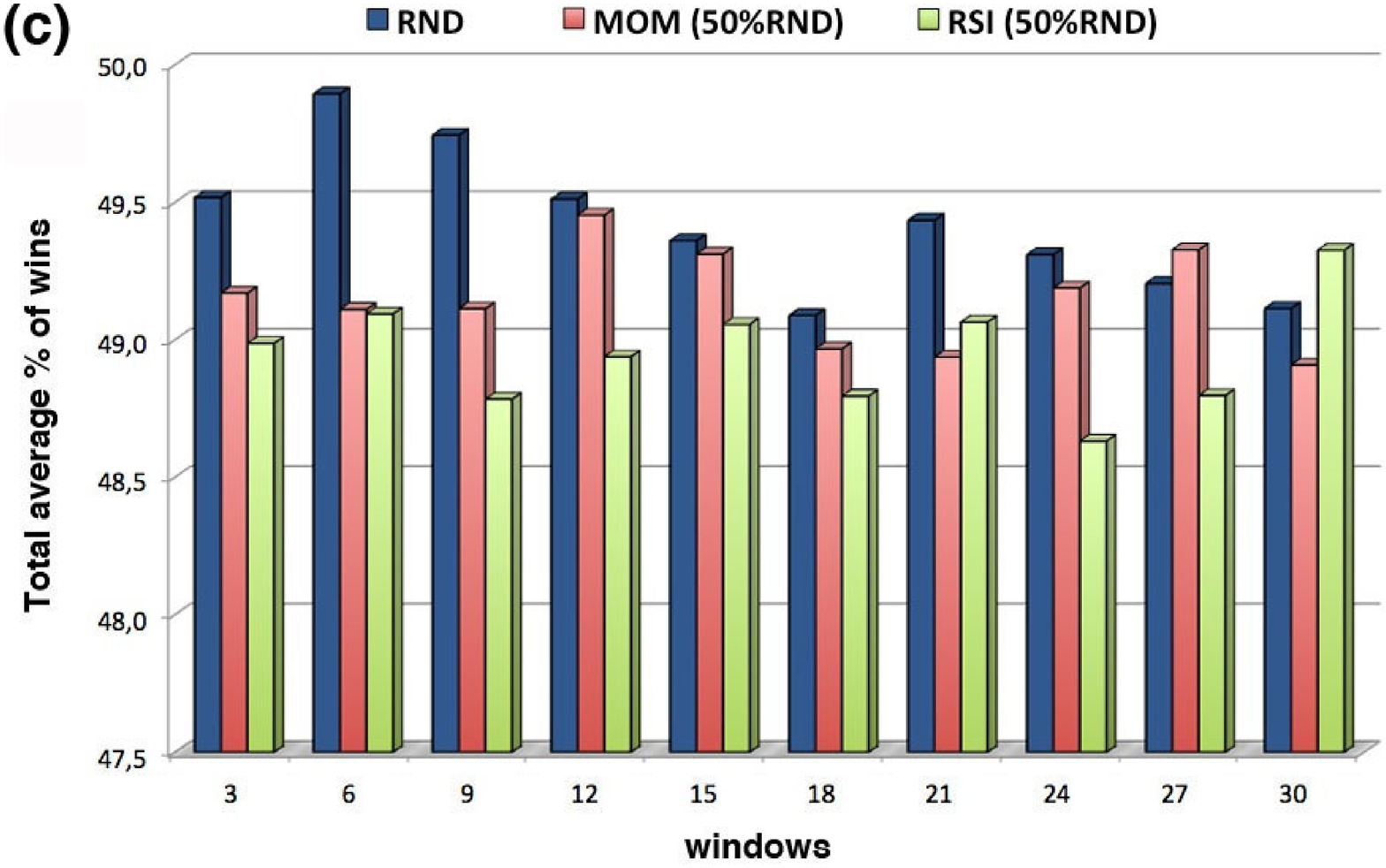}
\includegraphics[width=0.48\textwidth]{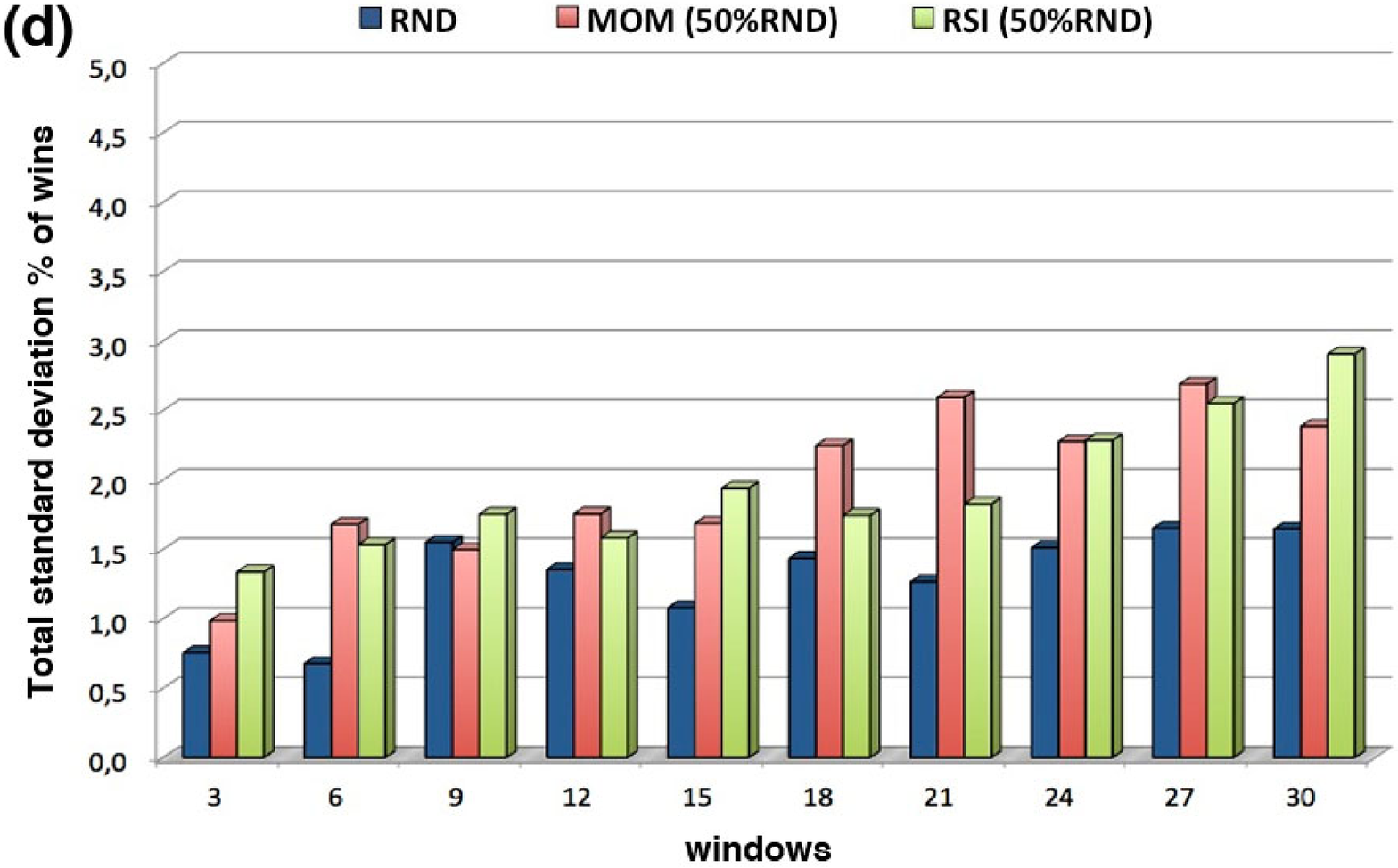}
\includegraphics[width=0.48\textwidth]{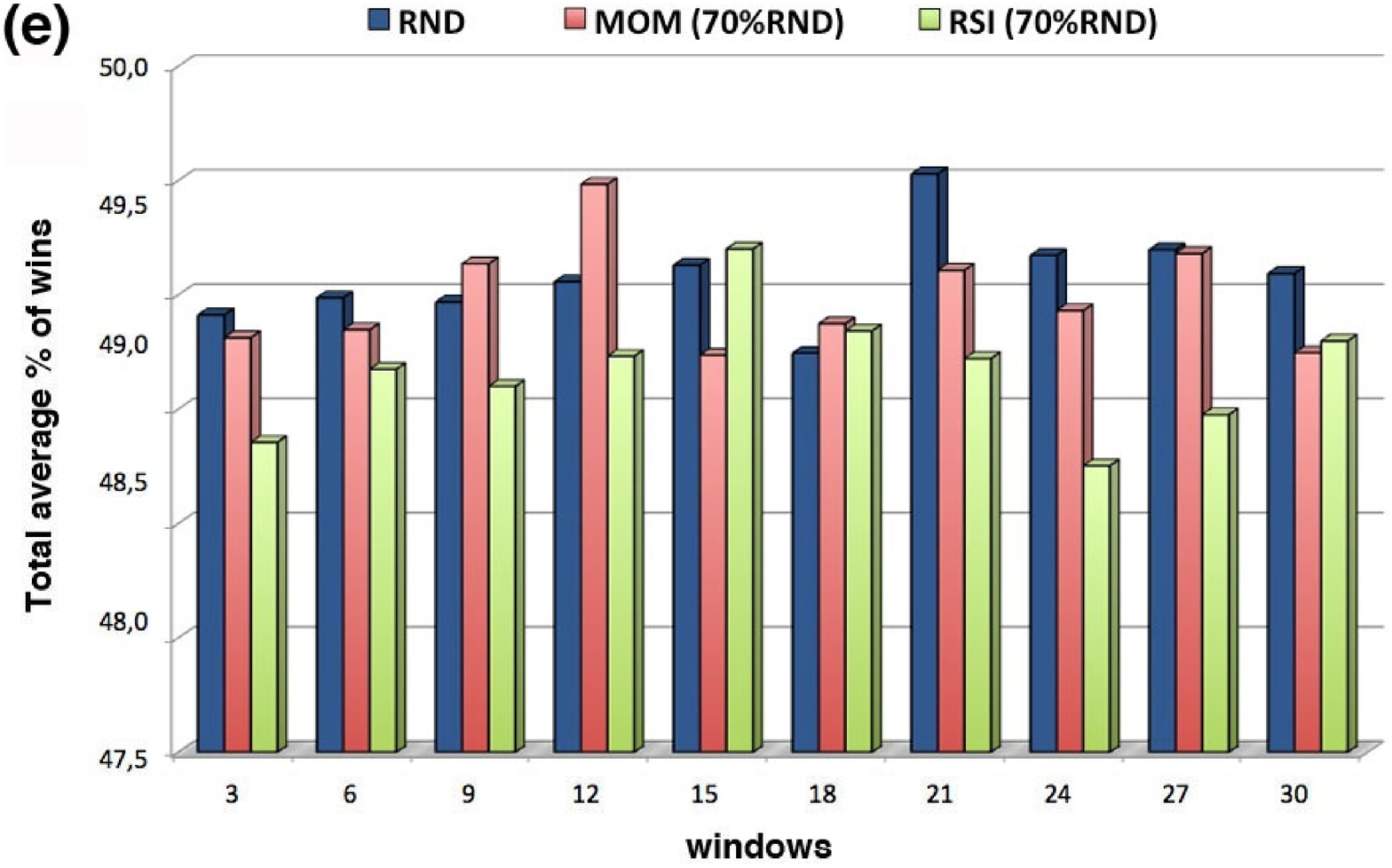}
\includegraphics[width=0.48\textwidth]{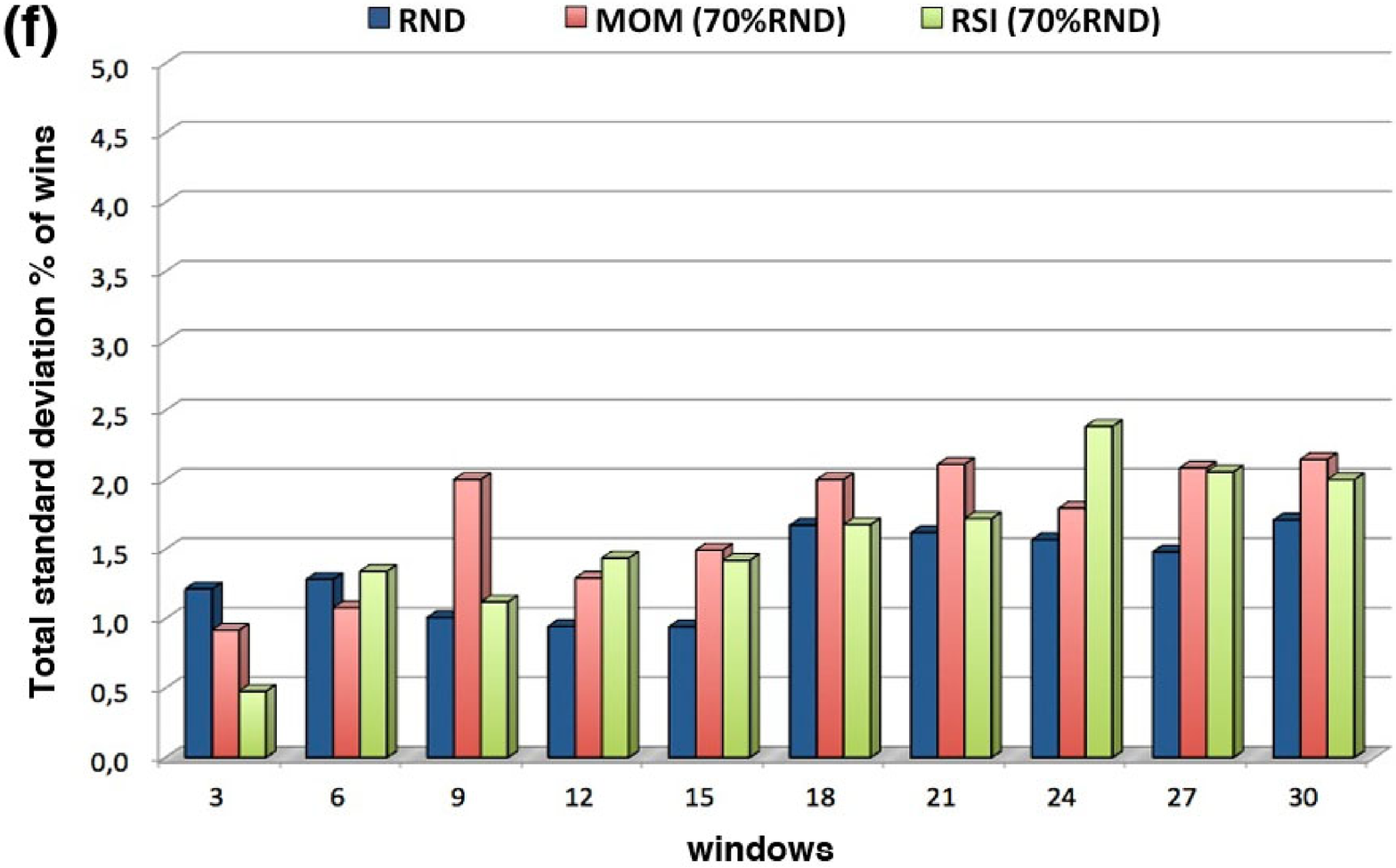}
\caption{Average percentage of wins (left column) with the correspondent standard deviations (right column) calculated over {\it all the windows} for the same configurations shown in Fig.6 and for the same three values of $P_{RND}$.}
\label{fig:7}       
\end{center}
\end{figure}
\\
In Fig.6 we present plots similar to those in Fig.4, but with an increasing percentage $P_{RND}$ of random predictions mixed with the two standard trading strategies, i.e. $20\%$ in panel (a), $50\%$ in panel (b), $70\%$ in panel (c), respectively. It immediately appears that the introduction of even a relatively small quantity of 'noise' (i.e. of random choices) in the MOM and RSI strategies improves their performance, in terms of both enhancing the average number of wins per window and stabilizing its fluctuations (i.e. reducing the trading risk) in each configuration (different columns). In this case, of course, the average over 10 runs performed inside each window (as in Fig.6) makes sense for all the three strategies (notice that we repeat here all the calculations also for the RND strategy, thus reinforcing the previous results).
We summarize these results in Fig.7, where we report a synthesis of both the averages and the standard deviations calculated over all the trading windows for the same configurations shown in Fig.6. It is evident that already for $P_{RND}=50\%$ the average gain of the MOM and RSI strategies becomes comparable with RND strategy (left column), as well as the corresponding fluctuations (right column). This further supports the analogy with the results found for the social systems presented in the first section, where the beneficial role of randomness could be appreciated even in moderate doses: actually, the same seems to happen also in financial trading, at least for the two standard strategies we considered in this paper.  
\\
The rationale behind the advantages to adopt some kind of random strategy for trading in financial markets, as suggested some years ago by the experiments of Wiseman and now corroborated by the results of our simulations, is twofold. On one hand, the intrinsic turbulent nature of financial markets makes any long term prediction about their behavior very difficult with the instruments of standard financial analysis, whose mathematical models are often based on unrealistic assumptions \cite{mandelbrot}. Such assumptions usually lead the traders to underestimate both the risks they face and the role of chance in the possible success of their strategies, at least until the next big market crash suddenly comes to reset their capital \cite{Taleb}. In this respect, the effectiveness of random strategies could be probably related to their stronger agreement with the turbulent and erratic essence of the financial markets.
On the other hand, last but not least, random strategies are also very cheap to implement: following them, everyone can invest in the stock market by himself/herself, reducing the  efforts to gather expensive  information and without resorting to costly financial consultants or to complicate trading rules. 

\section{Conclusions}

In this paper we have explored the beneficial  role of random strategies  in social and financial systems. We presented a short review of recent results  obtained  in the  managerial and political fields and then we focused our attention on financial markets. In particular, we numerically simulated the performance of three trading strategies (one completely random and two chosen among the most popular ones adopted by traders) applied to the FTSE UK All-Share index series, chosen here as a case study, in order to compare their predictive capability. Our results clearly indicate that (i) the standard strategies, with their algorithms based on the past history of the index, do not perform better than the purely random one, which, on the other hand, is also much less risky, and (ii) that the introduction of some degree of randomness in the same strategies significantly improves their performance. 
This means that random strategies offer also in the financial  field a better and costless alternative to the traditionally adopted strategies, minimizing both risk and volatility.
Of course one should investigate these findings on different stock price index series and consider a broader set of strategies in order to test the robustness and generality of these results. A study in this direction is in progress \cite{biondo}.

\end{document}